\documentclass[conference]{IEEEtran}
\usepackage{graphicx,psfrag,epsfig,epsf,latexsym,hhline,amsmath,amssymb,multirow}

\usepackage[usenames,dvipsnames]{pstricks}
\usepackage{pst-plot}
\usepackage{caption}
\usepackage{subcaption}
\usepackage{enumitem} 
\usepackage{amsmath} 
\usepackage{color}
\usepackage{hyperref}
\usepackage{graphicx}
\usepackage{amsthm}
\usepackage{algorithm}
\usepackage{algpseudocode}
\usepackage[noadjust]{cite}
\usepackage{blindtext}
\usepackage{etoolbox,tcolorbox}
\usepackage{stfloats}
\usepackage{tikz}
\usepackage{cite}
\usetikzlibrary{shapes.multipart}
\usetikzlibrary{chains,shapes.multipart}
\usetikzlibrary{shapes,calc,fit}
\usetikzlibrary{automata,positioning}
\graphicspath{ {figures/} }
\usepackage{pgfplots}
\pgfplotsset{compat=newest}
\usepackage{balance}
\usepgfplotslibrary{fillbetween}
\newcounter{cntr}
\usepackage{bbm}

\input{Jerry.def}
\definecolor{darkbrown}{rgb}{0.6, 0.26, 0.13}

\begin{document}
\title{Randomized Scheduling for PAoI Violation Guarantees in Periodic Multi-Source Systems}
\author{Kuan-Yu Lin$^\dag$, Wei-Lun Lu$^\dag$,  Yu-Pin Hsu$^*$, and Yu-Chih Huang$^\dag$\\
$^\dag$Institute of Communications Engineering, National Yang Ming Chiao Tung University, Hsinchu, Taiwan\\
$^*$Department of Communication Engineering, National Taipei University, New Taipei City, Taiwan\\
}

\maketitle

\begin{abstract}
We study peak Age of Information (PAoI) violation guarantee in a periodic multi-source status update system. The system is served by a shared base station, which requires scheduling. Our main contribution is a randomized scheduling framework that targets heterogeneous PAoI requirements. To that end, we derive numerically trackable upper bounds on the PAoI violation probability in two traffic regimes (long and short period) by leveraging the multivariate noncentral hypergeometric Wallenius distribution and the geometric distribution, respectively. Guided by these bounds, we design two low-complexity randomized scheduling schemes that meet diverse PAoI violation probability targets without the traffic assumption. Simulations validate the bounds and demonstrate feasible operation across a wide range of configurations.

\end{abstract}





\section{Introduction}\label{sec:intro}


Age of Information (AoI) has emerged as a critical performance metric in modern communication systems \cite{Yates21}, especially for applications requiring timely information updates, such as Internet of Things (IoT) networks \cite{Abd19}, industrial wireless networks (IWN) \cite{Li21}. In IoT networks, a central server must maintain fresh information to achieve its goals. For example, in autonomous vehicle and robotic networks, maintaining up-to-date information is critical for ensuring safe and effective operation in dynamic environments. Similarly, in IWNs, outdated updates may cause instability or production disruptions, thereby posing safety risks. 

Given the pivotal role of AoI in many modern applications, it has attracted extensive research attention in recent years. A large body of work has focused on analyzing the average AoI in both single-source and multi-source (requires scheduling) systems. For a single-source system, the prior \cite{Kaul12} showed that there exists an optimal update rate that minimizes the average AoI, highlighting the fundamental trade-off between update frequency and information freshness. Also, \cite{Sun17} revealed that persistently updating the status information is not always optimal for minimizing the average AoI. For a multi-source system, the prior \cite{Kadota18} proposed three low-complexity scheduling policies with provable average AoI performance. Later on, \cite{Hsu20} studied stochastic arrivals and characterized the structure of the optimal scheduling. Moreover, \cite{Wang24} investigated a scheduling design that accounts for the effects of not only AoI but also delay. For a comprehensive survey of recent advances in AoI research, the reader is referred to~\cite{Yates21}.

While minimizing the system’s \textit{average} AoI, as in~\cite{Kaul12, Sun17, Hsu20, Kadota18, Wang24}, typically enhances information timeliness, it does not directly provide statistical AoI guarantees. For systems with stringent timeliness requirements—such as autonomous control or real-time monitoring—ensuring \textit{statistical} AoI guarantees becomes essential. Accordingly, several studies have focused on characterizing AoI statistical behavior under single-source systems. For example, \cite{Seo19} analyzed the PAoI violation probability under D/G/1 queueing system. \cite{Inoue20} considered the scenario with infinite servers and examined the AoI distribution. \cite{Champati21} investigated multi-hop systems and derived upper bounds on the AoI violation probability.

However, the study of AoI statistical behavior in multi-source systems remains largely open, particularly regarding how scheduling impacts AoI statistics. A few example studies include \cite{Lin23, Li24}. \cite{Lin23} proposed a deterministic scheduling algorithm where users’ scheduling frequencies are proportional to their arrival rates. Large deviation theory was then applied to show that the PAoI violation probability decays exponentially under the proposed scheduling policy. Nevertheless, the approach in~\cite{Lin23} cannot fulfill all possible AoI violation requirements. In contrast,~\cite{Li24} proposed a feasible-optimal scheduling algorithm under the assumption of saturated traffic. Such an assumption may be impractical, as sensor data generation times and the scheduling times may not be synchronized, resulting in misalignment between data availability and transmission opportunities. Therefore, we aim to design a new scheduling algorithm that does not rely on this assumption. This presents an additional challenge, as the schedulable sets can vary over time.

Considering unsaturated traffic, our main contribution is the development of a randomized scheduling framework that provides tailored statistical AoI guarantees for each source according to its heterogeneous peak AoI (PAoI) requirements. To achieve this goal, we derive analytical upper bounds on the PAoI violation probability for a fixed scheduling distribution. In particular, we express these bounds under two traffic regimes (long and short sampling periods) by the multivariate noncentral hypergeometric Wallenius distribution and the geometric distribution, respectively. These expressions enable us to efficiently infer the scheduling distribution that meets a desired PAoI violation target. Based on these findings, we propose two randomized scheduling algorithms to meet diverse PAoI violation requirements without the two traffic assumptions. Simulation and numerical results demonstrate the practical effectiveness of the proposed scheduling designs across diverse system scenarios and validate both the accuracy and scalability of the derived analytical bounds.

\section{System Model and Problem Formulation} \label{sec:system model}
In this section, we first present the network model in Section~\ref{subsec:net_model}, then provide the definition of PAoI and a description of our problem in Section~\ref{subsec:AoI}.

\subsection{Network Model}\label{subsec:net_model}
We consider an information update system illustrated in Fig.~\ref{fig: illustrate_network}, where $n$ sources aim to update their respective statuses to a destination through a shared base station (BS). The sources generate new information updates (in the form of packets) simultaneously and periodically, resulting in a periodic packet arrival pattern at the BS. We denote the packet sampling period by $b$.

The BS maintains a queue for each source, following the single packet queueing discipline. i.e., at most one packet can stay in a queue. A packet in the queue is preempted upon the arrival of a new packet. We assume that the BS can transmit at most one packet at a time. A scheduling policy determines which queue to serve whenever the BS is available.

Due to channel uncertainties, we consider a stochastic transmission time for each packet. Let $V_i(k)$ represent the transmission time of the $k$-th updated packet from source $i$. We assume $V_i(k)$ is independent and identically distributed (i.i.d.) across different sources and packets, with its log moment generating function $\Lambda(\theta) = \log \mathbb{E} [e^{\theta V_i(k)}]$ for all $i$ that exists. Note that the transmission time can be either discrete or continuous.

\subsection{Peak Age of Information and Problem Formulation}\label{subsec:AoI}
For each source $i$, we use the pair $(i, k)$ to denote the $k$-th update received by the BS. Note that, due to the single-packet queuing discipline, not all packets arriving at the BS can be received by it.  Denote $S_i(k)$ as the arrival time of the $(i,k)$-th update from source $i$ and $D_i(k)$ as the departure time of the $(i,k)$-th update, which depends on the scheduling design. Then, PAoI at the destination upon receiving update $(i,k)$ is defined by \cite{costa14},
\begin{equation}\label{eqn:A(k)}
    A_i(k) = D_i(k) - S_i(k-1),
\end{equation}
which represents the maximum age reached before receiving update $(i, k)$. Specifically, it captures the time between the generation time of the previous update $(i,k-1)$ and the departure time of the current update $(i,k)$.

While most works focused on the long-term average AoI or PAoI, we consider the PAoI violation probability, as defined in Definition \ref{def:age_violation_probability}, to provide a statistical guarantee.
\begin{define}[PAoI violation probability]\label{def:age_violation_probability}
    The PAoI violation probability for update $(i,k)$ is defined as the probability that PAoI upon receiving update $(i, k)$ exceeds a given threshold $x>0$, which can be  expressed as,
    \begin{equation}
        \text{Pr}(A_i(k) \geq x).
    \end{equation}
\end{define}
Furthermore, we investigate the PAoI violation probability in a large-source regime. The asymptotic decay rate defined in Definition \ref{def:asymptotic decay rate}, characterizes how the PAoI violation probability exponent scales with the number of sources.
\begin{define}[Asymptotic decay rate]\label{def:asymptotic decay rate}
The asymptotic decay rate for update $(i,k)$ under a given threshold $x=n\cdot x'$ is defined by,
\begin{equation}\label{eqn:asymptotic_decay_rate}
    -\lim_{n \to \infty} \frac{1}{n} \log \text{Pr}(A_i(k) \geq nx').
\end{equation}
\end{define}

Next, Section~\ref{sec:randomized policy and age analysis} proposes a randomized scheduling framework and analyzes the PAoI violation probability under this framework. Building on these analytical results, Section~\ref{sec:design_algo} presents a scheduling distribution that achieves the required PAoI violation probabilities for heterogeneous sources.
\begin{figure}
\centering
\begin{tikzpicture}[>=latex]
\def\pos{0}
\draw [fill=red!60] (\pos,100pt) circle [radius=10pt] node (s1) {\small 1};
\draw [fill=red!60] (\pos,70pt) circle [radius=10pt] node (s2) {\small 2};
\draw [fill=red!60] (\pos,20pt) circle [radius=10pt] node (sn) {\small $n$};
\foreach \idx in {10, 15, 20}
    \filldraw [black] ([yshift=10+\idx pt]sn) circle (1pt);

\tikzset{
    queuei/.pic={
  \stepcounter{cntr}
        \node[outer sep=0pt,draw,rectangle split,rectangle split horizontal,minimum height=10pt,rectangle split parts=1] (queue-\thecntr) [pic actions] {};
        \draw
          (queue-\thecntr.north west) -- ++(-0pt,0)
          (queue-\thecntr.south west) -- ++(-0pt,0);
    },
}
\path
(\pos+60pt,100pt) pic[rectangle split parts=1, rectangle split part fill={gray!80}] {queuei=1}
(\pos+60pt,70pt) pic[rectangle split parts=1, rectangle split part fill={white}] {queuei=2}
(\pos+60pt,20pt) pic[rectangle split parts=2, rectangle split part fill={gray!80}] {queuei=n};
\path
(\pos+60pt,100pt) coordinate (Q1)
(\pos+60pt,70pt) coordinate (Q2)
(\pos+60pt,20pt) coordinate (Qn);
\foreach \idx in {10, 15, 20}
    \filldraw [black] ([yshift=10+\idx pt]Qn) circle (1pt);

\draw[line width=1pt, ->] ([xshift=+70pt]s1)--([xshift=-10pt]Q1);
\draw[line width=1pt, ->] ([xshift=+70pt]s2)--([xshift=-10pt]Q2);
\draw[line width=1pt, ->] ([xshift=+70pt]sn)--([xshift=-10pt]Qn);

\draw [fill=black!60!green!70]  (\pos+120pt,60pt) circle [radius=15pt] node{} coordinate (BS);

\draw[line width=1pt, ->] ([xshift=20pt]Q1)--([xshift=-19pt, yshift=3pt]BS);
\draw[line width=1pt, ->] ([xshift=20pt]Q2)--([xshift=-19pt, yshift=0pt]BS);
\draw[line width=1pt, ->] ([xshift=20pt]Qn)--([xshift=-19pt, yshift=-1pt]BS);


\draw [fill=green!20!blue!80] (\pos+190pt,60pt) circle [radius=15pt] node{} coordinate (Dn);

\draw[line width=1pt, ->] ([xshift=19pt]BS)--([xshift=-19]Dn);

\begin{scope}
\node [anchor=south] at ([yshift=5pt]s1.north) {\small Sources};
\node [anchor=south] at ([xshift=5pt, yshift=11pt]Q1.north) {\small Queues};
\node [anchor=south] at ([yshift=15pt]BS.north) {\small BS};
\node [anchor=south] at ([yshift=15pt]Dn.north) {\small Destination};
\end{scope}
\end{tikzpicture}
\caption{An illustration of the network model.}
\label{fig: illustrate_network}
\end{figure}

\section{Randomized Scheduling Policy \& PAoI Analysis}\label{sec:randomized policy and age analysis}
In this section, we begin by defining our randomized scheduling policy in Section~\ref{subsec:randomized_policy}. Then, we provide the PAoI analysis in \ref{subsec:age analysis}. After that, we derive an upper bound on the PAoI violation probability under the randomized policy for long and short sampling period cases in Section~\ref{subsec:age analysis long} and~\ref{subsec:age analysis short}, respectively.

\subsection{Randomized Scheduling Policy}\label{subsec:randomized_policy}

We propose a randomized scheduling policy as follows. Let $\boldsymbol{\mu}=\{\mu_1,\ldots,\mu_n \}$ be the scheduling weight vector, where $\mu_i$ represents the scheduling weight for source $i$, and it satisfies the condition $\sum_{j=1}^n \mu_j = 1$. Let $Q_i(t)$ indicate whether a packet is present in the queue for source $i$ at time $t$, where $Q_i(t) = 1$ if a packet is present and $Q_i(t) = 0$ otherwise. For each time~$t$ at which the BS is available, our randomized scheduling policy selects a non-empty queue, say for source $i$, to serve with probability $\frac{\mu_i}{\sum_{j=1}^n Q_{j}(t) \mu_{j}}$. In other words, the probability of selecting source $i$ is proportional to its weight relative to the sum of the weights of all sources with non-empty queues.


\subsection{PAoI Analysis}\label{subsec:age analysis}
\hspace{-10pt} We now analyze the PAoI violation probability for a given scheduling weight vector. The result will later be used in Section~\ref{sec:design_algo} to derive a weight vector that meets the PAoI violation requirements of each individual source. We start from expressing $D_{i}(k)$ in \eqref{eqn:A(k)} as follows:
\begin{equation}\label{eqn:D(k)}
    D_{i}(k) = S_{i}(k) + W_{i}(k) + V_{i}(k),
\end{equation}
where $W_{i}(k)$ denotes the waiting time of update $(i, k)$ in its queue. The waiting time of update $(i,k)$ can be further iteratively expressed by that of the previous update $(i, k-1)$, which is shown in the following Lemma,
\begin{lemma}\label{lemma:W_i(k)}
The waiting time of update $(i,k)$ can be expressed by that of the previous update $(i,k-1)$, i.e.,
\begin{align}\label{eqn:W(k)}
    W_{i}(k) = \ &W_{i}(k-1) + T_{i}(k-1) \nonumber\\
    &+ N_{i}(k-1) - (I_{i}(k-1) + 1 )b,
\end{align}
where $I_{i}(k-1)$ represents the number of preempted packets for source $i$ between update $(i,k-1)$ and $(i,k)$; $N_{i}(k-1)$ accounts for the total idle time of the BS from the start of transmitting update $(i,k-1)$ to the start of transmitting update $(i,k)$; $T_i(k-1)$ accounts for the total busy time of the BS during the same period.
\end{lemma}
\begin{IEEEproof}
    See Appendix~\ref{apx:proof_W_i(k)} in \cite{Lin25}.
\end{IEEEproof}

Before providing an upper bound on the PAoI violation probability, it is essential to note that handling the term $T_{i}(k-1)$ presents several challenges. First, unlike in our previous work \cite{Lin23}, the transmission scheduling here is not deterministic, so we cannot directly express the total transmission time as a fixed number of transmitted packets. Second, the current scheduling probability distribution depends on the number of packets remaining in the queues, which varies at different scheduling moments.

To tackle these challenges and offer theoretical insights, we analyze the PAoI violation probability in two distinct scenarios to guide scheduling design. The first case, called the long sampling period case, examines scenarios where the sources generate new information updates infrequently. Specifically, this case assumes that no packets arrive at the BS before all previously generated (simultaneous) packets have been served. The second case, called the short sampling period case, considers the opposite scenario, where the sampling period is so short that the BS is always busy. Later, in Section~\ref{sec:design_algo}, we will demonstrate how our analysis for the two traffic regimes behaves under more general traffic conditions.
\subsection{PAoI Analysis for Long Sampling Period Case}\label{subsec:age analysis long}
For the long sampling period case, we first provide an upper bound on PAoI in the following Lemma,
\begin{lemma}\label{lemma:peak age long}
    For the long sampling period case, PAoI of update $(i,k)$ can be upper bounded by,
    \begin{equation}
        A_i(k) \leq b + T'_{i}(k-1) + V_i(k),
    \end{equation}
    where $T'_i(k-1)$ accounts for the total busy time of the BS from the arrival of update $(i,k)$ to the start of transmitting update $(i,k)$.
\end{lemma}
\begin{IEEEproof}
    See Appendix~\ref{apx:proof UB PAoI long sampling} in \cite{Lin25}
\end{IEEEproof}
After providing the upper bound of PAoI, we introduce the multivariate Wallenius noncentral hypergeometric distribution. This distribution characterizes the outcomes in a biased item selection process without replacement, where each item has a distinct selection weight. See \cite{Fog08} for more details. This distribution naturally arises when modeling our randomized scheduling policy, as each source is selected according to its weight as well.

When the BS makes $\ell$ transmissions before transmitting update $(i,k)$, we use $y_{i,\ell,j}$ to denote the number of packets scheduled for source~$j$ during these $\ell$ transmissions. Let $\mathbf{y}_{i,\ell} = ( y_{i,\ell,1}, \dots, y_{i,\ell,n} )$, $\mathbbm{1}=(1,\ldots,1)$, and $d_{\mathbf{y}_{i,\ell}}=\sum_{j=1}^n \mu_j (1-y_{i,\ell,j})$. Then given a scheduling weight vector $\boldsymbol{\mu}$, the probability that the BS makes $\ell$ transmission which follows by $\mathbf{y}_{i,\ell}$ before transmitting update $(i,k)$ can be expressed using the multivariate Wallenius noncentral hypergeometric distribution $g(\mathbf{y}_{i,\ell},n,\mathbbm{1},\boldsymbol{\mu})$ \cite{Fog08} with notations adapted to our setting:
\begin{align}\label{eqn:wallenius_function}
    \int_0^1 \prod_{j=1}^{n} (1-t^{\mu_i/d_{\mathbf{y}_{i,\ell}}})^{y_{i,\ell,j}}dt, 
\end{align}
Based on these results, we can derive an upper bound for the PAoI violation probability under the long sampling period case,
\begin{theorem}\label{thm:UB_long_sampling_delay_case}
For the long sampling period case, given a scheduling weight vector $\boldsymbol{\mu}$, the PAoI violation probability for update $(i,k)$ under the randomized scheduling policy is upper bounded by,
\begin{align}
    \exp\left( -\sup_{\theta>0} \left\{ \theta x - \theta b - n\!\max_{0\leq \ell \leq n-1}f_i(\ell,n, \boldsymbol{\mu}) \right\} \right), \label{eqn:age violation prob long sampling delay case}
\end{align}
\end{theorem}
\!\!\!\!where,
\begin{align}
    &f_i(\ell,n, \boldsymbol{\mu}) = (\ell+2) \Lambda(\theta) \nonumber\\
    &+ \log \left( \sum_{\mathbf{y}_{i,\ell}\in S_{i,\ell}} g(\mathbf{y}_{i,\ell}, n, \mathbbm{1}, \boldsymbol{\mu}) \frac{\mu_i}{1-\sum_{j\in \mathbf{y}_{i,\ell}} \mu_j} \right),
\end{align} 
and $S_{i,\ell}$ is the set consisting of all possible $\mathbf{y}_{i,\ell}$.
\begin{IEEEproof}
    See Appendix~\ref{apx:proof_UB_long_sampling_delay_case} in \cite{Lin25}.
\end{IEEEproof}
The above theorem can also allow us to examine the asymptotic decay rate when the number of sources is large. We provide the asymptotic decay rate under the randomized scheduling policy for the long sampling period in the following corollary. Moreover, we will validate the result with numerical studies in Section \ref{sec:discussion_simulation}.
\begin{corollary}\label{coro:asym_decay_rate_rand_policy_long_sampling_delay}
    For the long sampling period case, given a scheduling weight vector $\boldsymbol{\mu}$, if the sampling period $b=n b'$ and the threshold $x=nx'$ also scale with $n$, then the asymptotic decay rate for update $(i,k)$ under the randomized scheduling policy is lower bounded by,
    \begin{align}
        \sup_{\theta>0} \left\{ \theta x' - \theta b' - f_i(\infty,\infty,\boldsymbol{\mu}) \right\}.\label{eqn:asym_decay_rate_rand_policy_long_sampling_delay}
    \end{align}
\end{corollary}
\begin{IEEEproof}
    See Appendix~\ref{apx:proof_asymptotic_decay_rate_rand_policy_long_sampling_delay} in \cite{Lin25}.
\end{IEEEproof}

\subsection{PAoI Analysis for Short Sampling Period Case}\label{subsec:age analysis short}
For the short sampling period case, we start by providing an upper bound on PAoI in the following Lemma,
\begin{lemma}\label{lemma:peak age short}
    For the short sampling period case, PAoI of update $(i, k)$ can be upper bounded by,		\begin{equation}\label{eqn:peak age short}
    A_{i}(k) \leq b + T_{i}(k-1) + V_{i}(k).
    \end{equation}
\end{lemma}
\begin{IEEEproof}
    See Appendix~\ref{apx:proof UB PAoI short sampling} in \cite{Lin25}
\end{IEEEproof}
After providing an upper bound for PAoI, we provide an upper bound for the PAoI violation probability in the following, 
\begin{theorem}\label{thm:UB_short_sampling_delay_case}
For the short sampling period case, given a scheduling weight vector $\boldsymbol{\mu}$, if $\Lambda(\theta) < \log \left( \frac{1}{1 - \mu_i} \right)$, the PAoI violation probability for update $(i,k)$ under the randomized scheduling policy is upper bounded by,
    \begin{align}\label{eqn:age_violation_prob_short_sampling_delay_case}
        \Pr\left(A_{i}(k) \geq x\right) \leq \inf_{\theta>0} \left\{ e^{-\theta (x-b)} \frac{e^{\Lambda(\theta)}\mu_i}{1 - e^{\Lambda(\theta)}(1 - \mu_i)} \right\}.
    \end{align}
\end{theorem}
\begin{IEEEproof}
    See Appendix~\ref{apx:proof_UB_short_sampling_delay_case} in \cite{Lin25}.
\end{IEEEproof}
Note that, unlike Theorem~\ref{thm:UB_short_sampling_delay_case}, Theorem 1 does not require any upper bound condition on the transmission time. This is because, in the long sampling period case, all currently arrived packets are served before the next arrivals, which implicitly bounds the transmission times. Next, we also derive the asymptotic decay rate in the following, 
\begin{corollary}\label{coro:asym_decay_rate_rand_policy_short_sampling_delay}
    For the short sampling period case, given a scheduling weight vector $\boldsymbol{\mu}$, if $\Lambda(\theta) < \log \left( \frac{1}{1 - \mu_i} \right)$, $b=n b'$, and $x=nx'$, then the asymptotic decay rate of the packet $(i,k)$ under the randomized scheduling policy is lower bounded by,
    \begin{equation}\label{eqn:asym_decay_rate_rand_policy_short_sampling_delay}
        \sup_{\theta>0} \left\{ \theta x' - \theta b' \right\}.
    \end{equation}
\end{corollary}
\begin{IEEEproof}
    See Appendix~\ref{apx:proof_asymptotic_decay_rate_rand_policy_short_sampling_delay} in \cite{Lin25}.
\end{IEEEproof}

\section{Randomized Scheduling Weight Design}\label{sec:design_algo}
After providing the theoretical finding in the previous section, we are now ready to propose algorithms to determine a scheduling weight vector $\boldsymbol{\mu}$ that achieves a desired PAoI violation probability vector $
\boldsymbol{\epsilon}_i=\{ \epsilon_1, \cdots, \epsilon_n\}$, where $\epsilon_i$ represents the source $i$. Specifically, we propose a scheduling weight design based on Theorem~\ref{thm:UB_long_sampling_delay_case}, and another one based on Theorem~\ref{thm:UB_short_sampling_delay_case}.

\subsection{Randomized-L Scheduling Weight Design}
We start with the use of Theorem ~\ref{thm:UB_long_sampling_delay_case}. To apply the analytical bound in \eqref{eqn:age violation prob long sampling delay case} of Theorem~\ref{thm:UB_long_sampling_delay_case}, two additional challenges arise. The first challenge is to efficiently search for the $\max_{0\leq \ell \leq n-1}f_i(\ell,n, \boldsymbol{\mu})$ in \eqref{eqn:age violation prob long sampling delay case}. To address this, we propose the following Lemma~\ref{lemma:exponent term converge to n} to reduce the computational complexity by exploiting its asymptotic property. 
\begin{lemma}\label{lemma:exponent term converge to n}
     When $n$ is sufficiently large, for all $0\leq j \leq n-1$, we have
\begin{equation}
    f_i(n-1,n,\boldsymbol{\mu}) \geq f_i(j,n,\boldsymbol{\mu}).
\end{equation}
\end{lemma}
\begin{IEEEproof}
    We first prove that when $n$ is large, for all $1 \leq \ell \leq n-1$, we have $f_i(\ell,n,\boldsymbol{\mu}) \leq f_i(\ell+1,n,\boldsymbol{\mu})$. Since it implies that $f_i(\ell,n,\boldsymbol{\mu})$ is monotonically increasing in $\ell$, we complete the proof. See Appendix \ref{apx:proof_exp_term_increasing} in \cite{Lin25} for details.
\end{IEEEproof}
Then we can upper bound $\max_{0\leq \ell \leq n-1}f_i(\ell,n, \boldsymbol{\mu})$ in the following,
\begin{equation}
     \max_{0\leq \ell \leq n-1} f_i(\ell,n,\boldsymbol{\mu}) \leq f_i(n-1,n,\boldsymbol{\mu}).\label{eqn:approx maxf}
\end{equation}
To calculate $f_i(n-1,n,\boldsymbol{\mu})$, the integration in \eqref{eqn:wallenius_function} for the term $g(\mathbf{y}_{i,n-1},n,\mathbbm{1},\boldsymbol{\mu})$ poses a second challenge, as the integration has no closed form. To address this, we employ a Taylor-series-based approximation to $g(\mathbf{y}_{i,n-1},n,\mathbbm{1},\boldsymbol{\mu})$ as introduced in \cite{Fog08}.
\begin{align}
    \hspace{-10pt}
    g(\mathbf{y}_{i,n-1},n,\mathbbm{1},\boldsymbol{\mu}) \approx \Phi(\tau,\mathbf{y}_{i,n-1}, \boldsymbol{\mu}) \sqrt{\frac{-2\pi}{\psi(\tau,\mathbf{y}_{i,n-1},\boldsymbol{\mu})}}, \label{eqn:numerical form of g}
\end{align}
where 
\begin{align}
    &\Phi(\tau, \mathbf{y}_{i,n-1},\boldsymbol{\mu}) = rd\tau^{rd-1}\prod_{j=1}^n(1-\tau^{r\mu_j})^{y_{i,\ell,j}},\\
    &\psi(\tau,\mathbf{y}_{i,n-1},\boldsymbol{\mu}) = -\frac{rd-1}{\tau^2} \nonumber\\
    &\hspace{-17pt}-\sum_{j=1}^n y_{i,\ell,j} r \mu_j \frac{(r \mu_j - 1)\tau^{r \mu_j-2}(1-\tau^{r\mu_j})+r\mu_j\tau^{2r\mu_j-2}}{(1-\tau^{r\mu_j})^2},
\end{align}
and $\tau>0, r>0$ are some constant values. The choice of $\tau$ and $r$ can be found in \cite{Fog08}. Next, plugging \eqref{eqn:approx maxf} and \eqref{eqn:numerical form of g} into \eqref{eqn:age violation prob long sampling delay case} yield,
\begin{align}
    &\text{Pr} \left( A_{i}{(k)} \geq x \right) \leq \exp\left( \theta_i^* x_i - \theta_i^* b - (n+2) \Lambda(\theta_i^*) \vphantom{\sqrt{\frac{-2\pi}{\psi(\tau,\mathbf{y}_{i,\ell})}}} \right. \nonumber\\
    &\ \left. + \log\left( \Phi(\tau,\mathbf{y}_{i,n-1},\boldsymbol{\mu}) \sqrt{\frac{-2\pi}{\psi(\tau,\mathbf{y}_{i,n-1},\boldsymbol{\mu})}} \right) \right),
\end{align}
where $\theta^* = \argmin_{\theta>0} \left\{ \theta x - \theta b - nf_i(n-1,n,\boldsymbol{\mu})\right\}$.

We are now ready to propose a numerically tractable algorithm, called Randomized-L, to determine the scheduling weights that satisfy a specified age violation constraint; that is, to guarantee the following equation for all $i$,
\begin{align}
    &\exp\left( \theta_i^* x_i - \theta_i^* b - (n+2)\Lambda(\theta_i^*) \vphantom{\sqrt{\frac{-2\pi}{\psi(\tau,\mathbf{y}_{i,\ell})}}} \right. \nonumber\\
    &\ \left. + \log\left( \Phi(\tau,\mathbf{y}_{i,n-1},\boldsymbol{\mu}) \sqrt{\frac{-2\pi}{\psi(\tau,\mathbf{y}_{i,n-1},\boldsymbol{\mu})}} \right) \right) \leq \epsilon_i.\label{eqn:age_violation_constraint_long_sampling_delay}
\end{align}

\begin{figure*}[t]
    \centering
    \begin{subfigure}[t]{0.25\textwidth}
        \centering
        \resizebox{\textwidth}{!}{
%
%
\definecolor{mycolor1}{rgb}{0.85000,0.32500,0.09800}%
\definecolor{mycolor3}{rgb}{0.00000,0.44700,0.74100}%
\definecolor{mycolor2}{rgb}{0.46667,0.67451,0.18824}%

\begin{tikzpicture}

\begin{axis}[%
width=4.521in,
height=3.566in,
at={(0.758in,0.481in)},
scale only axis,
xmode=log,
xmin=1e-06,
xmax=0.1,
xminorticks=true,
xlabel style={font=\fontsize{20}{16}\selectfont, color=white!15!black},
xlabel={$\epsilon_2$},
ymin=0,
ymax=1,
ylabel style={font=\fontsize{20}{16}\selectfont, color=white!15!black},
ylabel={$\mu_1'/\mu_2'$},
axis background/.style={fill=white},
xmajorgrids,
xminorgrids,
ymajorgrids,
legend style={at={(0.032,0.723)}, anchor=south west, legend cell align=left, align=left,font=\fontsize{16}{16}\selectfont, draw=white!15!black}
]
\addplot [color=mycolor1, line width=1.5pt,fill=mycolor1!20, mark=asterisk, mark options={solid, mycolor1}, mark size = 4]
  table[row sep=crcr]{%
0.1	1\\
0.1	1\\
0.0666666666666667	1\\
0.0444444444444444	1\\
0.0296296296296296	1\\
0.0197530864197531	1\\
0.0131687242798354	1\\
0.00877914951989026	1\\
0.00585276634659351	1\\
0.00390184423106234	1\\
0.00260122948737489	1\\
0.00173415299158326	1\\
0.00115610199438884	1\\
0.000770734662925894	1\\
0.000513823108617262	1\\
0.000342548739078175	1\\
0.000228365826052117	1\\
0.000152243884034744	1\\
0.00010149592268983	0.694915254237288\\
6.76639484598864e-05	0.449275362318841\\
4.51092989732576e-05	0.351351351351351\\
3.00728659821717e-05	0.219512195121951\\
2.00485773214478e-05	0.136363636363636\\
1.33657182142986e-05	0.0526315789473684\\
8.91047880953237e-06	inf\\
5.94031920635491e-06	inf\\
3.96021280423661e-06	inf\\
2.64014186949107e-06	inf\\
1.76009457966071e-06	inf\\
1.17339638644048e-06	inf\\
7.82264257626984e-07	inf\\
5.21509505084656e-07	inf\\
3.47673003389771e-07	inf\\
2.31782002259847e-07	inf\\
1.54521334839898e-07	inf\\
1.03014223226599e-07	inf\\
} -- (axis cs:1.33657182142986e-05,0) -- (axis cs:0.1,0) -- cycle;
\addlegendentry{Opt randomized}

\addplot [color=mycolor3, line width=1.5pt,fill=mycolor3!20, mark=o, mark options={solid, mycolor3}, mark size = 4]
  table[row sep=crcr]{%
0.1	1\\
0.1	1\\
0.0666666666666667	1\\
0.0444444444444444	1\\
0.0296296296296296	1\\
0.0197530864197531	1\\
0.0131687242798354	1\\
0.00877914951989026	1\\
0.00585276634659351	1\\
0.00390184423106234	1\\
0.00260122948737489	1\\
0.00173415299158326	1\\
0.00115610199438884	1\\
0.000770734662925894	0.515151515151515\\
0.000513823108617262	0.298701298701299\\
0.000342548739078175	0.176470588235294\\
0.000228365826052117	0.111111111111111\\
0.000152243884034744	0.0638297872340426\\
0.00010149592268983	0.0416666666666667\\
6.76639484598864e-05	0.0309278350515464\\
4.51092989732576e-05	0.0204081632653061\\
3.00728659821717e-05	0.0101010101010101\\
2.00485773214478e-05	inf\\
1.33657182142986e-05	inf\\
8.91047880953237e-06	inf\\
5.94031920635491e-06	inf\\
3.96021280423661e-06	inf\\
2.64014186949107e-06	inf\\
1.76009457966071e-06	inf\\
1.17339638644048e-06	inf\\
7.82264257626984e-07	inf\\
5.21509505084656e-07	inf\\
3.47673003389771e-07	inf\\
2.31782002259847e-07	inf\\
1.54521334839898e-07	inf\\
1.03014223226599e-07	inf\\
} -- (axis cs:3.00728659821717e-05,0) -- (axis cs:0.1,0) -- cycle;
\addlegendentry{Randomized-L}

\addplot [color=mycolor2, line width=1.5pt,fill=mycolor2!20, mark=x, mark options={solid, mycolor2}, mark size = 4]
  table[row sep=crcr]{%
0.1	1\\
0.1	1\\
0.0666666666666667	1\\
0.0444444444444444	1\\
0.0296296296296296	1\\
0.0197530864197531	0.851851851851852\\
0.0131687242798354	0.724137931034483\\
0.00877914951989026	0.639344262295082\\
0.00585276634659351	0.538461538461539\\
0.00390184423106234	0.449275362318841\\
0.00260122948737489	0.388888888888889\\
0.00173415299158326	0.315789473684211\\
0.00115610199438884	0.265822784810127\\
0.000770734662925894	inf\\
0.000513823108617262	inf\\
0.000342548739078175	inf\\
0.000228365826052117	inf\\
0.000152243884034744	inf\\
0.00010149592268983	inf\\
6.76639484598864e-05	inf\\
4.51092989732576e-05	inf\\
3.00728659821717e-05	inf\\
2.00485773214478e-05	inf\\
1.33657182142986e-05	inf\\
8.91047880953237e-06	inf\\
5.94031920635491e-06	inf\\
3.96021280423661e-06	inf\\
2.64014186949107e-06	inf\\
1.76009457966071e-06	inf\\
1.17339638644048e-06	inf\\
7.82264257626984e-07	inf\\
5.21509505084656e-07	inf\\
3.47673003389771e-07	inf\\
2.31782002259847e-07	inf\\
1.54521334839898e-07	inf\\
1.03014223226599e-07	inf\\
} -- (axis cs:0.00115610199438884,0) -- (axis cs:0.1,0) -- cycle;
\addlegendentry{Randomized-S}

\end{axis}

\begin{axis}[%
width=5.833in,
height=4.375in,
at={(0in,0in)},
scale only axis,
xmin=0,
xmax=1,
ymin=0,
ymax=1,
axis line style={draw=none},
ticks=none,
axis x line*=bottom,
axis y line*=left
]
\end{axis}
\end{tikzpicture}%
}
        \caption{Scenario 1 $(\mathbb{E}[V_i(k)] = 1.5)$}
        \label{fig:feasible_region_rand_long_sampling_delay}
    \end{subfigure}
    \hfill
    \begin{subfigure}[t]{0.25\textwidth}
        \centering
        \resizebox{\textwidth}{!}{
%
%
\definecolor{mycolor1}{rgb}{0.85000,0.32500,0.09800}%
\definecolor{mycolor3}{rgb}{0.00000,0.44700,0.74100}%
\definecolor{mycolor2}{rgb}{0.46667,0.67451,0.18824}%

\begin{tikzpicture}

\begin{axis}[%
width=4.521in,
height=3.566in,
at={(0.758in,0.481in)},
scale only axis,
unbounded coords=jump,
xmode=log,
xmin=1e-06,
xmax=0.1,
xminorticks=true,
xlabel style={font=\fontsize{20}{16}\selectfont, color=white!15!black},
xlabel={$\epsilon_2$},
ymin=0,
ymax=1,
ylabel style={font=\fontsize{20}{16}\selectfont, color=white!15!black},
ylabel={$\mu_1'/\mu_2'$},
axis background/.style={fill=white},
xmajorgrids,
xminorgrids,
ymajorgrids,
legend style={at={(0.032,0.723)}, anchor=south west, legend cell align=left, align=left,font=\fontsize{16}{16}\selectfont, draw=white!15!black}
]

\addplot [color=mycolor1, line width=1.5pt,fill=mycolor1!20, mark=asterisk, mark options={solid, mycolor1}, mark size = 4]
  table[row sep=crcr]{%
0.1	1\\
0.1	1\\
0.0666666666666667	1\\
0.0444444444444444	1\\
0.0296296296296296	1\\
0.0197530864197531	1\\
0.0131687242798354	1\\
0.00877914951989026	1\\
0.00585276634659351	1\\
0.00390184423106234	1\\
0.00260122948737489	1\\
0.00173415299158326	1\\
0.00115610199438884	1\\
0.000770734662925894	1\\
0.000513823108617262	1\\
0.000342548739078175	0.886792452830189\\
0.000228365826052117	0.785714285714286\\
0.000152243884034744	0.694915254237288\\
0.00010149592268983	0.587301587301587\\
6.76639484598864e-05	0.492537313432836\\
4.51092989732576e-05	0.428571428571429\\
3.00728659821717e-05	0.351351351351351\\
2.00485773214478e-05	0.265822784810127\\
1.33657182142986e-05	0.234567901234568\\
8.91047880953237e-06	0.19047619047619\\
5.94031920635491e-06	0.136363636363636\\
3.96021280423661e-06	0.0638297872340426\\
2.64014186949107e-06	inf\\
1.76009457966071e-06	inf\\
1.17339638644048e-06	inf\\
7.82264257626984e-07	inf\\
5.21509505084656e-07	inf\\
3.47673003389771e-07	inf\\
2.31782002259847e-07	inf\\
1.54521334839898e-07	inf\\
1.03014223226599e-07	inf\\
} -- (axis cs:3.00728659821718e-06,0) -- (axis cs:0.1,0) -- cycle;
\addlegendentry{Opt randomized}

\addplot [color=mycolor3, line width=1.5pt,fill=mycolor3!20, mark=o, mark options={solid, mycolor3}, mark size = 4]
  table[row sep=crcr]{%
0.1	1\\
0.1	1\\
0.0666666666666667	1\\
0.0444444444444444	1\\
0.0296296296296296	1\\
0.0197530864197531	1\\
0.0131687242798354	1\\
0.00877914951989026	1\\
0.00585276634659351	1\\
0.00390184423106234	1\\
0.00260122948737489	1\\
0.00173415299158326	1\\
0.00115610199438884	1\\
0.000770734662925894	1\\
0.000513823108617262	1\\
0.000342548739078175	1\\
0.000228365826052117	1\\
0.000152243884034744	1\\
0.00010149592268983	0.96078431372549\\
6.76639484598864e-05	0.492537313432836\\
4.51092989732576e-05	0.282051282051282\\
3.00728659821717e-05	0.176470588235294\\
2.00485773214478e-05	0.111111111111111\\
1.33657182142986e-05	0.0638297872340426\\
8.91047880953237e-06	0.0416666666666667\\
5.94031920635491e-06	0.0204081632653061\\
3.96021280423661e-06	0.0101010101010101\\
2.64014186949107e-06	0.0101010101010101\\
1.76009457966071e-06	inf\\
1.17339638644048e-06	inf\\
7.82264257626984e-07	inf\\
5.21509505084656e-07	inf\\
3.47673003389771e-07	inf\\
2.31782002259847e-07	inf\\
1.54521334839898e-07	inf\\
1.03014223226599e-07	inf\\
} -- (axis cs:2.00485773214478e-06,0) -- (axis cs:0.1,0) -- cycle;
\addlegendentry{Randomized-L}

\addplot [color=mycolor2, line width=1.5pt,fill=mycolor2!20, mark=o, mark options={solid, mycolor2}, mark size = 4]
  table[row sep=crcr]{%
0.1	1\\
0.1	1\\
0.0666666666666667	1\\
0.0444444444444444	1\\
0.0296296296296296	1\\
0.0197530864197531	1\\
0.0131687242798354	1\\
0.00877914951989026	1\\
0.00585276634659351	0.886792452830189\\
0.00390184423106234	0.785714285714286\\
0.00260122948737489	0.694915254237288\\
0.00173415299158326	0.612903225806452\\
0.00115610199438884	0.538461538461539\\
0.000770734662925894	0.470588235294118\\
0.000513823108617262	0.408450704225352\\
0.000342548739078175	0.351351351351351\\
0.000228365826052117	0.298701298701299\\
0.000152243884034744	0.265822784810127\\
0.00010149592268983	0.219512195121951\\
6.76639484598864e-05	inf\\
4.51092989732576e-05	inf\\
3.00728659821717e-05	inf\\
2.00485773214478e-05	inf\\
1.33657182142986e-05	inf\\
8.91047880953237e-06	inf\\
5.94031920635491e-06	inf\\
3.96021280423661e-06	inf\\
2.64014186949107e-06	inf\\
1.76009457966071e-06	inf\\
1.17339638644048e-06	inf\\
7.82264257626984e-07	inf\\
5.21509505084656e-07	inf\\
3.47673003389771e-07	inf\\
2.31782002259847e-07	inf\\
1.54521334839898e-07	inf\\
1.03014223226599e-07	inf\\
} -- (axis cs:0.00010149592268983,0) -- (axis cs:0.1,0) -- cycle;
\addlegendentry{Randomized-S}

\end{axis}

\begin{axis}[%
width=5.833in,
height=4.375in,
at={(0in,0in)},
scale only axis,
xmin=0,
xmax=1,
ymin=0,
ymax=1,
axis line style={draw=none},
ticks=none,
axis x line*=bottom,
axis y line*=left
]
\end{axis}
\end{tikzpicture}%
}
        \caption{Scenario 2 $(\mathbb{E}[V_i(k)] = 8)$}
        \label{fig:feasible_region_rand_short_sampling_delay}
    \end{subfigure}
    \hfill
    \begin{subfigure}[t]{0.25\textwidth}
        \centering
        \resizebox{\textwidth}{!}{
%
%
\definecolor{mycolor1}{rgb}{0.85000,0.32500,0.09800}%
\definecolor{mycolor3}{rgb}{0.00000,0.44700,0.74100}%
\definecolor{mycolor2}{rgb}{0.46667,0.67451,0.18824}%

\begin{tikzpicture}

\begin{axis}[%
width=4.521in,
height=3.566in,
at={(0.758in,0.481in)},
scale only axis,
xmode=log,
xmin=1e-07,
xmax=0.1,
xminorticks=true,
xlabel style={font=\fontsize{20}{16}\selectfont, color=white!15!black},
xlabel={$\epsilon_2$},
ymin=0,
ymax=1,
ylabel style={font=\fontsize{20}{16}\selectfont, color=white!15!black},
ylabel={$\mu_1'/\mu_2'$},
axis background/.style={fill=white},
xmajorgrids,
xminorgrids,
ymajorgrids,
legend style={at={(0.032,0.723)}, anchor=south west, legend cell align=left, align=left,font=\fontsize{16}{16}\selectfont, draw=white!15!black}
]
\addplot [color=mycolor1, line width=1.5pt,fill=mycolor1!20, mark=asterisk, mark options={solid, mycolor1}, mark size = 4]
  table[row sep=crcr]{%
0.1	1\\
0.1	1\\
0.0666666666666667	1\\
0.0444444444444444	1\\
0.0296296296296296	1\\
0.0197530864197531	1\\
0.0131687242798354	1\\
0.00877914951989026	1\\
0.00585276634659351	1\\
0.00390184423106234	1\\
0.00260122948737489	1\\
0.00173415299158326	1\\
0.00115610199438884	1\\
0.000770734662925894	1\\
0.000513823108617262	1\\
0.000342548739078175	1\\
0.000228365826052117	1\\
0.000152243884034744	0.923076923076923\\
0.00010149592268983	0.724137931034483\\
6.76639484598864e-05	0.449275362318841\\
4.51092989732576e-05	0.388888888888889\\
3.00728659821717e-05	0.282051282051282\\
2.00485773214478e-05	0.136363636363636\\
1.33657182142986e-05	0.0869565217391304\\
8.91047880953237e-06	0.0752688172043011\\
5.94031920635491e-06	0.0101010101010101\\
3.96021280423661e-06	inf\\
2.64014186949107e-06	inf\\
1.76009457966071e-06	inf\\
1.17339638644048e-06	inf\\
7.82264257626984e-07	inf\\
5.21509505084656e-07	inf\\
3.47673003389771e-07	inf\\
2.31782002259847e-07	inf\\
1.54521334839898e-07	inf\\
1.03014223226599e-07	inf\\
} -- (axis cs:5.94031920635491e-06,0) -- (axis cs:0.1,0) -- cycle;
\addlegendentry{Opt randomized}

\addplot [color=mycolor3, line width=1.5pt,fill=mycolor3!20, mark=o, mark options={solid, mycolor3}, mark size = 4]
  table[row sep=crcr]{%
0.1	1\\
0.1	1\\
0.0666666666666667	1\\
0.0444444444444444	1\\
0.0296296296296296	1\\
0.0197530864197531	1\\
0.0131687242798354	1\\
0.00877914951989026	1\\
0.00585276634659351	1\\
0.00390184423106234	1\\
0.00260122948737489	1\\
0.00173415299158326	1\\
0.00115610199438884	1\\
0.000770734662925894	1\\
0.000513823108617262	0.818181818181818\\
0.000342548739078175	0.428571428571429\\
0.000228365826052117	0.25\\
0.000152243884034744	0.149425287356322\\
0.00010149592268983	0.0989010989010989\\
6.76639484598864e-05	0.0638297872340426\\
4.51092989732576e-05	0.0416666666666667\\
3.00728659821717e-05	0.0204081632653061\\
2.00485773214478e-05	0.0101010101010101\\
1.33657182142986e-05	0.0101010101010101\\
8.91047880953237e-06	inf\\
5.94031920635491e-06	inf\\
3.96021280423661e-06	inf\\
2.64014186949107e-06	inf\\
1.76009457966071e-06	inf\\
1.17339638644048e-06	inf\\
7.82264257626984e-07	inf\\
5.21509505084656e-07	inf\\
3.47673003389771e-07	inf\\
2.31782002259847e-07	inf\\
1.54521334839898e-07	inf\\
1.03014223226599e-07	inf\\
} -- (axis cs:1.33657182142986e-05,0) -- (axis cs:0.1,0) -- cycle;
\addlegendentry{Randomized-L}

\addplot [color=mycolor2, line width=1.5pt,fill=mycolor2!20, mark=x, mark options={solid, mycolor2}, mark size = 4]
  table[row sep=crcr]{%
0.1	1\\
0.1	1\\
0.0666666666666667	1\\
0.0444444444444444	1\\
0.0296296296296296	1\\
0.0197530864197531	0.96078431372549\\
0.0131687242798354	0.851851851851852\\
0.00877914951989026	0.724137931034483\\
0.00585276634659351	0.639344262295082\\
0.00390184423106234	0.538461538461539\\
0.00260122948737489	0.470588235294118\\
0.00173415299158326	0.408450704225352\\
0.00115610199438884	0.333333333333333\\
0.000770734662925894	0.282051282051282\\
0.000513823108617262	inf\\
0.000342548739078175	inf\\
0.000228365826052117	inf\\
0.000152243884034744	inf\\
0.00010149592268983	inf\\
6.76639484598864e-05	inf\\
4.51092989732576e-05	inf\\
3.00728659821717e-05	inf\\
2.00485773214478e-05	inf\\
1.33657182142986e-05	inf\\
8.91047880953237e-06	inf\\
5.94031920635491e-06	inf\\
3.96021280423661e-06	inf\\
2.64014186949107e-06	inf\\
1.76009457966071e-06	inf\\
1.17339638644048e-06	inf\\
7.82264257626984e-07	inf\\
5.21509505084656e-07	inf\\
3.47673003389771e-07	inf\\
2.31782002259847e-07	inf\\
1.54521334839898e-07	inf\\
1.03014223226599e-07	inf\\
} -- (axis cs:0.000770734662925894,0) -- (axis cs:0.1,0) -- cycle;
\addlegendentry{Randomized-S}

\end{axis}

\begin{axis}[%
width=5.833in,
height=4.375in,
at={(0in,0in)},
scale only axis,
xmin=0,
xmax=1,
ymin=0,
ymax=1,
axis line style={draw=none},
ticks=none,
axis x line*=bottom,
axis y line*=left
]
\end{axis}
\end{tikzpicture}%
}
        \caption{Scenario 3 $(\mathbb{E}[V_i(k)] = 3)$}
        \label{fig:feasible_region_rand_typical}
    \end{subfigure}

    \caption{Achievable region of our randomized policy under different scenarios}
    \label{fig:feasible_region_rand}
\end{figure*}

\noindent\fbox{%
\parbox{0.97\linewidth}{%
\underline{Randomized-L scheduling algorithm.}\\ 
Given $\epsilon_1 \leq \epsilon_2 \leq \ldots \leq \epsilon_n$. We identify a scheduling weight vector $\boldsymbol{\mu}$ through an $n$-iteration procedure: for each iteration $i$, we determine $\mu_i$ by searching for the minimum weight for source~$i$ that meets \eqref{eqn:age_violation_constraint_long_sampling_delay} subject to the constraint that  $\mu_j=(1-\sum_{k=1}^{i-1}\mu_k-\mu_i)/(n-i)$ remains equal for all $j > i$, where $\mu_k$ for all $k<i$ have been determined in the prior iterations. 


}}
\vspace{2pt}

The intuition underlying the Randomized-L scheduling weight design is as follows. We iteratively determine the scheduling weights from the most AoI-critical source to the least AoI-critical one, i.e., from source~1 to~$n$. At the beginning of each iteration~$i$, the weights $\mu_j$ for all $j < i$ have already been determined in the previous iterations, and $\mu_i$ is the variable to be determined in the current iteration. To mitigate the computational complexity caused by the curse of dimensionality in the remaining variables $\mu_i, \ldots, \mu_n$, we divide the sources whose weights are yet to be determined into two subsets: one consisting of source~$i$, and the other including all sources $j > i$. Leveraging the idea of mean-field analysis \cite{mitzenmacher2002power}, we treat the sources in the latter subset symmetrically by fixing $\mu_j$ for all $j > i$ to a common value (representing the mean behavior of that subset). We then search for only a single parameter $\mu_i$ that satisfies the constraint in~\eqref{eqn:age_violation_constraint_long_sampling_delay}. Here, we identify the minimum weight that guarantees the PAoI violation probability constraint for source~$i$, while preserving the remaining weights for allocation to the other sources in later iterations.

To search for the value in each iteration~$i$, we divide the range of the weight, between 0 and~1, into $\Delta$ sub-intervals. We then start from the minimum value and incrementally increase it until we first find a $\mu_i$ that satisfies the constraint in~\eqref{eqn:age_violation_constraint_long_sampling_delay}. Thus, for each iteration, at most $\Delta$ computations are required. Considering a total of $n$ iterations, the overall computational complexity is $\mathcal{O}(n\Delta)$.
\vspace{-1pt}
\subsection{Randomized-S Scheduling Weight Design}
Next, we also use Theorem 2 to propose a scheduling weight design, called Randomized-S, to determine a scheduling weight vector that satisfies a specified age-violation constraint; that is, to guarantee the following equation for all $i$, 
\begin{align}
    \eqref{eqn:age_violation_prob_short_sampling_delay_case} &\leq \epsilon_i 
    \ \Rightarrow \  \mu_i \leq \frac{\epsilon_i \cdot \left(1 - e^{\Lambda(\theta_{i}^*)} \right)}
    {e^{\Lambda(\theta_{i}^*)}\left(e^{-\theta_{i}^* x_i + \theta_{i}^* b} - \epsilon_i\right)}, \label{eqn:mu_i upper bound}
\end{align}
where $\theta^* = \argmin_{\theta>0} \left\{ e^{-\theta (x-b)} \frac{e^{\Lambda(\theta)}\mu_i}{1 - e^{\Lambda(\theta)}(1 - \mu_i)} \right\}$. Moreover, applying the upper bound in \eqref{eqn:age_violation_prob_short_sampling_delay_case} requires an additional condition $\Lambda(\theta_i^*) < \log \left( \frac{1}{1 - \mu_i} \right)$, which yields a lower bound on the scheduling weight: $\mu_i \geq 1- e^{-\Lambda(\theta_{i}^*)}$.
Combining these two bounds, we have,
\begin{equation}\label{eqn:age_violation_constraint_short_sampling_delay}
    1- e^{-\Lambda(\theta_{i}^*) } \leq \mu_i \leq \frac{\epsilon_i \cdot \left(1 - e^{\Lambda(\theta_{i}^*)} \right)}
    {e^{\Lambda(\theta_{i}^*)}\left(e^{-\theta_{i}^* x_i + \theta_{i}^* b} - \epsilon_i\right)}. 
\end{equation}

Our randomized-S scheduling algorithm is expressed in the following,
\vspace{2pt}

\noindent\fbox{%
\parbox{0.97\linewidth}{%
\underline{Randomized-S scheduling algorithm.}\\ 
Given $\epsilon_1,\ldots, \epsilon_n$, we identify a minimum value of $\mu_i$ that meets \eqref{eqn:age_violation_constraint_short_sampling_delay}.
}}


\begin{figure*}[t]
    \centering
    \begin{subfigure}[t]{0.25\textwidth}
        \centering
        \resizebox{\textwidth}{!}{
%
%
\definecolor{mycolor1}{rgb}{0.00000,0.44706,0.74118}%
\definecolor{mycolor2}{rgb}{0.46667,0.67451,0.18824}%
\definecolor{mycolor3}{rgb}{0.70000,0.44706,0.74118}%
\definecolor{mycolor4}{rgb}{0.858,0.188,0.278}%

\begin{tikzpicture}

\begin{axis}[%
width=4.521in,
height=3.555in,
at={(0.758in,0.492in)},
scale only axis,
xmin=5,
xmax=50,
xlabel style={font=\fontsize{17}{16}\selectfont, color=white!15!black},
xlabel={Total number of sources n},
ymode=log,
ymin=1e-6,
ymax=1,
yminorticks=true,
ylabel style={font=\fontsize{17}{16}\selectfont, color=white!15!black},
ylabel={PAoI violation probability},
axis background/.style={fill=white},
xmajorgrids,
ymajorgrids,
yminorgrids,
legend style={
  at={(0.221,0.509)},
  anchor=south west,
  legend cell align=left,
  align=left,
  draw=white!15!black,
  font=\fontsize{16}{16}\selectfont
}
]


\addplot [color=mycolor3, line width=1.6pt, mark size=5.0pt, mark=triangle, mark options={solid, rotate=90, mycolor3}]
  table[row sep=crcr]{%
6	0.354908439565015\\
12	0.0538770302236934\\
18	0.00830798399814921\\
24	0.00125103111539009\\
30	0.000202040930825340\\
36	2.69845425973406e-05\\
42	4.19513002871055e-06\\
48	5.20116295786377e-07\\
};
\addlegendentry{Numerical randomized-L group 1}

\addplot [color=mycolor3, line width=1.6pt, mark size=4.0pt, mark=square, mark options={solid, rotate=90, mycolor3}]
  table[row sep=crcr]{%
6	0.686844312623478\\
12	0.686844312623478\\
18	0.686844312623478\\
24	0.686844312623478\\
30	0.686844312623478\\
36	0.686844312623478\\
42	0.686844312623478\\
48	0.686844312623478\\
};
\addlegendentry{Numerical randomized-S group 1}

\addplot [color=mycolor3, dashed, line width=1.6pt, mark size=4.0pt, mark=o, mark options={solid, rotate=90, mycolor3}]
  table[row sep=crcr]{%
6	0.00866687233425884\\
12	0.000652333333333333\\
18	6.38888888888889e-05\\
24	5.45833333333333e-06\\
30	4.66666666666667e-07\\
36	2.77777777777778e-08\\
42	0\\
48	0\\
};
\addlegendentry{Simulation randomized group 1}

\addplot [color=mycolor1, line width=1.6pt, mark size=5.0pt, mark=triangle, mark options={solid, rotate=90, mycolor1}]
  table[row sep=crcr]{%
6	0.0241302851883261\\
12	0.00130098823704856\\
18	9.44705468103196e-05\\
24	8.05796880179682e-06\\
30	7.58334023618986e-07\\
36	7.72292091894537e-08\\
42	8.26648339776371e-09\\
48	9.28224222264696e-10\\
};
\addlegendentry{Numerical randomized-L group 2}

\addplot [color=mycolor1, line width=1.6pt, mark size=4.0pt, mark=square, mark options={solid, rotate=90, mycolor1}]
  table[row sep=crcr]{%
6	1.0\\
12	1.0\\
18	1.0\\
24	1.0\\
30	1.0\\
36	1.0\\
42	1.0\\
48	1.0\\
};
\addlegendentry{Numerical randomized-S group 2}

\addplot [color=mycolor1, dashed, line width=1.6pt, mark size=4.0pt, mark=o, mark options={solid, rotate=90, mycolor1}]
  table[row sep=crcr]{%
6	0.00156633411650039\\
12	5.32499955625004e-05\\
18	2.49999986111112e-06\\
24	8.33333298611113e-08\\
30	0\\
36	0\\
42	0\\
48	0\\
};
\addlegendentry{Simulation randomized group 2}



\end{axis}
\end{tikzpicture}%
}
        \caption{Scenario 1 $(\mathbb{E}[V_i(k)] = 1.5)$}
        \label{fig:long_sample_delay_scale_by_n_group1}
    \end{subfigure}
    \hfill
    \begin{subfigure}[t]{0.25\textwidth}
        \centering
        \resizebox{\textwidth}{!}{
%
%
\definecolor{mycolor1}{rgb}{0.00000,0.44706,0.74118}%
\definecolor{mycolor2}{rgb}{0.46667,0.67451,0.18824}%
\definecolor{mycolor3}{rgb}{0.70000,0.44706,0.74118}%
\definecolor{mycolor4}{rgb}{0.858,0.188,0.278}%
\begin{tikzpicture}

\begin{axis}[%
width=4.521in,
height=3.548in,
at={(0.758in,0.499in)},
scale only axis,
xmin=5,
xmax=50,
xlabel style={font=\fontsize{17}{16}\selectfont, color=white!15!black},
xlabel={Total number of sources n},
ymode=log,
ymin=1e-06,
ymax=1,
yminorticks=true,
ylabel style={font=\fontsize{17}{16}\selectfont, color=white!15!black},
ylabel={PAoI violation probability},
axis background/.style={fill=white},
xmajorgrids,
ymajorgrids,
yminorgrids,
legend style={
  at={(0.221,0.219)},
  anchor=south west,
  legend cell align=left,
  align=left,
  draw=white!15!black,
  font=\fontsize{16}{16}\selectfont
}
]

\addplot [color=mycolor3, line width=1.6pt, mark size=5.0pt, mark=triangle, mark options={solid, rotate=90, mycolor3}]
  table[row sep=crcr]{%
6	0.0674051292987638\\
12	0.00194337623436826\\
18	5.69148386918139e-05\\
24	1.62770084429709e-06\\
30	0\\
36	0\\
42	0\\
48	0\\
};
\addlegendentry{Numerical randomized-L group 1}

\addplot [color=mycolor3, line width=1.6pt, mark size=4.0pt, mark=square, mark options={solid, rotate=90, mycolor3}]
  table[row sep=crcr]{%
6	0.955375081104748\\
12	0.95537510250205\\
18	0.955375081104751\\
24	0.955375102502044\\
30	0.955375114739975\\
36	0.955375310288692\\
42	0.955375310288728\\
48	0.95537534721382\\
};
\addlegendentry{Numerical randomized-S group 1}

\addplot [color=mycolor3, dashed, line width=1.6pt, mark size=4.0pt, mark=o, mark options={solid, rotate=90, mycolor3}]
  table[row sep=crcr]{%
6	0.110713540554544\\
12	0.126377294553817\\
18	0.133475540112762\\
24	0.137049363066468\\
30	0.139053207936105\\
36	0.140089657857961\\
42	0.14115205388809\\
48	0.141689979004306\\
};
\addlegendentry{Simulation randomized group 1}

\addplot [color=mycolor1, line width=1.6pt, mark size=5.0pt, mark=triangle, mark options={solid, rotate=90, mycolor1}]
  table[row sep=crcr]{%
6	0.00458288620898577\\
12	4.69274124905454e-05\\
18	6.47181787306595e-07\\
24	0\\
30	0\\
36	0\\
42	0\\
48	0\\
};
\addlegendentry{Numerical randomized-L group 2}

\addplot [color=mycolor1, line width=1.6pt, mark size=4.0pt, mark=square, mark options={solid, rotate=90, mycolor1}]
  table[row sep=crcr]{%
6	0.50366827427827\\
12	0.503668277098609\\
18	0.503668278710948\\
24	0.503668277098607\\
30	0.503668278710948\\
36	0.503668278710949\\
42	0.503668277098606\\
48	0.50366827709861\\
};
\addlegendentry{Numerical randomized-S group 2}

\addplot [color=mycolor1, dashed, line width=1.6pt, mark size=4.0pt, mark=o, mark options={solid, rotate=90, mycolor1}]
  table[row sep=crcr]{%
6	0.0512082627889748\\
12	0.0332338085602696\\
18	0.0285014084954765\\
24	0.0263916035973495\\
30	0.0252397222859237\\
36	0.02448340292088\\
42	0.0239828807831451\\
48	0.0235628657754156\\
};
\addlegendentry{Simulation randomized group 2}

\end{axis}
\end{tikzpicture}%
}
        \caption{Scenario 2 $(\mathbb{E}[V_i(k)] = 8)$}
        \label{fig:short_sample_delay_scale_by_n_group1}
    \end{subfigure}
    \hfill
    \begin{subfigure}[t]{0.25\textwidth}
        \centering
        \resizebox{\textwidth}{!}{
%
%
\definecolor{mycolor1}{rgb}{0.00000,0.44706,0.74118}%
\definecolor{mycolor2}{rgb}{0.46667,0.67451,0.18824}%
\definecolor{mycolor3}{rgb}{0.70000,0.44706,0.74118}%
\definecolor{mycolor4}{rgb}{0.858,0.188,0.278}%
\begin{tikzpicture}

\begin{axis}[%
width=4.521in,
height=3.548in,
at={(0.758in,0.499in)},
scale only axis,
xmin=5,
xmax=50,
xlabel style={font=\fontsize{17}{16}\selectfont, color=white!15!black},
xlabel={Total number of sources n},
ymode=log,
ymin=1e-6,
ymax=1,
yminorticks=true,
ylabel style={font=\fontsize{17}{16}\selectfont, color=white!15!black},
ylabel={PAoI violation probability},
axis background/.style={fill=white},
xmajorgrids,
ymajorgrids,
yminorgrids,
legend style={
  at={(0.221,0.519)},
  anchor=south west,
  legend cell align=left,
  align=left,
  draw=white!15!black,
  font=\fontsize{16}{16}\selectfont
}
]

\addplot [color=mycolor3, line width=1.6pt, mark size=4.0pt, mark=square, mark options={solid, rotate=90, mycolor3}]
  table[row sep=crcr]{%
6	0.503668274278269\\
12	0.503668274278269\\
18	0.503668274278269\\
24	0.503668274278269\\
30	0.503668274278269\\
36	0.503668274278269\\
42	0.503668274278269\\
48	0.503668274278269\\
};
\addlegendentry{Numerical randomized-S group 1}

\addplot [color=mycolor3, line width=1.6pt, mark size=5.0pt, mark=triangle, mark options={solid, rotate=90, mycolor3}]
  table[row sep=crcr]{%
6	0.878249350892097\\
12	0.329918500839159\\
18	0.125892594348152\\
24	0.0469109420325592\\
30	0.0187476533244193\\
36	0.00619618141848893\\
42	0.0023837241759691\\
48	0.000731328580898394\\
};
\addlegendentry{Numerical randomized-L group 1}

\addplot [color=mycolor3, dashed, line width=1.6pt, mark size=4.0pt, mark=o, mark options={solid, rotate=90, mycolor3}]
  table[row sep=crcr]{%
6	0.032299681931805\\
12	0.00623303130554388\\
18	0.00149364159895818\\
24	0.000422175015674237\\
30	0.000131542857036183\\
36	3.46117398910525e-05\\
42	1.28572346945335e-05\\
48	5.20834830733472e-06\\
};
\addlegendentry{Simulation randomized group 1}

\addplot [color=mycolor1, line width=1.6pt, mark size=5.0pt, mark=triangle, mark options={solid, rotate=90, mycolor1}]
  table[row sep=crcr]{%
6	0.0597123227879915\\
12	0.00796666198924387\\
18	0.00143153167243572\\
24	0.000302156279497009\\
30	7.03668475533355e-05\\
36	1.77333445330086e-05\\
42	4.69711694050954e-06\\
48	1.3051636888208e-06\\
};
\addlegendentry{Numerical randomized-L group 2}

\addplot [color=mycolor1, line width=1.6pt, mark size=4.0pt, mark=square, mark options={solid, rotate=90, mycolor1}]
  table[row sep=crcr]{%
6	1.0\\
12	1.0\\
18	1.0\\
24	1.0\\
30	1.0\\
36	1.0\\
42	1.0\\
48	1.0\\
};
\addlegendentry{Numerical randomized-S group 2}

\addplot [color=mycolor1, dashed, line width=1.6pt, mark size=4.0pt, mark=o, mark options={solid, rotate=90, mycolor1}]
  table[row sep=crcr]{%
6	0.00685642932334386\\
12	0.000609675811803844\\
18	8.78896505991941e-05\\
24	1.5333413833756e-05\\
30	4.26667463112598e-06\\
36	7.77778166666861e-07\\
42	1.42857149659864e-07\\
48	4.16666673611111e-08\\
};
\addlegendentry{Simulation randomized group 2}

\end{axis}
\end{tikzpicture}%
}
        \caption{Scenario 3 $(\mathbb{E}[V_i(k)] = 3)$}
        \label{fig:long_sample_delay_scale_by_n_group1_extreme_case}
    \end{subfigure}
   \vspace{-2pt}
    \caption{PAoI violation probability under different scenarios}
    \vspace{-5pt}\label{fig:PAoI_violation_group1}
\end{figure*}

\section{Simulation Results}\label{sec:discussion_simulation}
In this section, we validate our scheduling weight design across various scenarios (beyond the short and long sampling period assumptions) to confirm its feasibility. We consider a system involving two groups of sources. Both Group 1 and Group 2 each have $n/2$ sources, with the PAoI violation probability constraints being $\epsilon_1$ and $\epsilon_2$, respectively. For the transmission, we set $\mu_i = \mu_1'/(n/2)$ for each source~$i$ in Group~1, and $\mu_i = \mu_2'/(n/2)$ for each source~$i$ in Group~2, where $\mu_1'$ and $\mu_2'$ are yet to be determined. We simulate the transmission time using an exponential distribution. 

In Fig.~\ref{fig:feasible_region_rand}, we show the PAoI violation probabilities that can be achieved by the proposed algorithms. We consider a system with $n = 18$ and $b = 90$. To evaluate the effectiveness of our scheduling schemes, we examine three representative scenarios with $\mathbb{E}[V_i(k)]=1.5$, 8, and 3 for Scenarios~1,~2, and~3, shown in Figs.~\ref{fig:feasible_region_rand_long_sampling_delay},~\ref{fig:feasible_region_rand_short_sampling_delay}, and~\ref{fig:feasible_region_rand_typical}, respectively. By fixing $\epsilon_1 = 0.1$, we identify $\mu_1'$ and $\mu_2'$ (the $y$-axis shows their resulting ratio) that meet different $\epsilon_2$ values (on the $x$-axis), using our proposed algorithms, where the sub-interval is set to $\Delta = 0.01$. Moreover, we compare our results with the optimized scheduling weights obtained by enumerating all possible parameter combinations to verify whether they satisfy the given PAoI constraint.
\vspace{-2pt}
In Fig.~\ref{fig:feasible_region_rand_long_sampling_delay}, we observe that Randomized-L achieves the same feasible region as the optimal randomized scheduling policy for $\epsilon_2 \ge 10^{-3}$, and it always provides a feasible solution for $\epsilon_2 \ge 5 \times 10^{-5}$. In contrast, Randomized-S yields feasible solutions only when $\epsilon_2 \ge 10^{-3}$. In Fig.~\ref{fig:feasible_region_rand_short_sampling_delay}, the average transmission time is much larger than that in Scenario~1. Although Randomized-L provides a broader achievable region, it slightly overestimates the true feasible region due to the approximation used for the Wallenius distribution. Fig.~\ref{fig:feasible_region_rand_typical} illustrates an intermediate scenario between Scenario~1 and Scenario~2. We can observe that Randomized-L also outperforms Randomized-S. Overall, these results demonstrate that Randomized-L consistently outperforms Randomized-S, regardless of the relationship between the sampling period and the transmission time.

In Fig.~\ref{fig:PAoI_violation_group1}, we validate the derived asymptotic decay rates (in Corollaries~\ref{coro:asym_decay_rate_rand_policy_long_sampling_delay} and \ref{coro:asym_decay_rate_rand_policy_short_sampling_delay}) by comparing them with computer simulations, where each subfigure uses the same average transmission times as in Fig.~\ref{fig:feasible_region_rand}. We set $b' = 5$, $(x_1', x_2') = (8,25)$, and $(\mu_1', \mu_2') = (0.2, 0.8)$. The number of sources $n$ varies from~6 to~48 (shown on the $x$-axis).
In Fig.~\ref{fig:long_sample_delay_scale_by_n_group1}, we observe that the asymptotic decay rate for the long sampling period case in Corollary~\ref{coro:asym_decay_rate_rand_policy_long_sampling_delay} accurately captures the slope, with only a constant gap. In contrast, the asymptotic decay rate for the short sampling period case in Corollary~\ref{coro:asym_decay_rate_rand_policy_short_sampling_delay} is overly loose. For Scenario~2, shown in Fig.~\ref{fig:short_sample_delay_scale_by_n_group1}, the asymptotic decay rate for the short sampling period case captures the exact slope, while that for the long sampling period case exhibits a similar overestimation behavior as observed in Fig.~\ref{fig:feasible_region_rand_short_sampling_delay}, indicating that our analytical upper bound is conservative. Finally, in Fig.~\ref{fig:long_sample_delay_scale_by_n_group1_extreme_case}, the results show that the asymptotic decay rate for the long sampling period case again captures the exact slope.

\section{Conclusion}\label{sec:conclusion}
We investigated the PAoI violation probability in a periodic multi-source status update system. In this system, we proposed two computationally efficient randomized scheduling schemes that enable the base station to meet the PAoI requirements of individual sources. In particular, our designs provide probabilistic guarantees under the short and long sampling period assumptions. Simulation results further validate the effectiveness of the proposed algorithms across various scenarios. Future work includes tightening the bounds in Theorem~\ref{thm:UB_long_sampling_delay_case} to address the overestimation issue observed in Fig.~\ref{fig:feasible_region_rand}, and extending the framework to handle heterogeneous sampling periods and other realistic system variations.
\bibliographystyle{IEEEtran}
\bibliography{bib_5g.bib}
\onecolumn
\begin{appendices}

\section{Proof of Lemma ~\ref{lemma:W_i(k)}}\label{apx:proof_W_i(k)}
\begin{IEEEproof}
    We begin the proof by considering the case where no packets are preempted, i.e., $I_i(k-1) = 0$. The waiting time of update $(i,k)$ can be expressed as the sum of the previous waiting time $W_i(k-1)$, the total busy time of the BS during this period, and the total idle time $N_i(k-1)$ from the start of transmitting update $(i, k-1)$ to the start of transmitting update $(i,k)$. We have,
    \begin{align}
        W_i(k) = W_i(k-1) + T_i(k-1) + N_i(k-1) - b.
    \end{align}
    Next, when $I_i(k-1)>0$, the term $W_i(k-1) + T_i(k-1)$ must exceed $b$, resulting in $I_i(k-1)$ packets being preempted. This yields the expression in \eqref{eqn:W(k)}, completing the proof.
\end{IEEEproof}

\section{Proof of Lemma~\ref{lemma:peak age long}}\label{apx:proof UB PAoI long sampling}
    \begin{IEEEproof}
    We start from \eqref{eqn:A(k)} and substitute $D_{i}(k)$ and $W_{i}(k)$ with \eqref{eqn:D(k)} and \eqref{eqn:W(k)}, respectively,
    \begin{align}
        &A_{i}(k) = D_{i}(k) - S_{i}(k-1)\nonumber\\
        &\overset{(a)}{=} S_{i}(k) + W_{i}(k-1) + T_{i}(k-1) + N_{i}(k-1) - (I_{i}(k-1)+1) b + V_{i}(k) -S_{i}(k-1) \nonumber\\
        &\overset{(b)}{=} W_{i}(k-1) + T_{i}(k-1) + N_{i}(k-1) + V_{i}(k) \nonumber\\
        &\overset{(c)}{\leq} b + T'_i(k-1) + V_i(k). \nonumber
    \end{align}
    where (a) applies \eqref{eqn:D(k)} and \eqref{eqn:W(k)}; (b) and (c) are both due to the assumption that no packets arrive at the BS before transmitting all previously generated packets. Specifically, (b) appies $I_i(k-1)=0$ and $S_i(k) - S_i(k-1) = b$; (c) applies $W_i(k-1) + \hat{T}-_i(k-1) + N_i(k-1) = b$, where $\hat{T}_i(k-1)$ accounts for the total busy time of the BS from the start of transmitting update $(i,k-1)$ to the end of transmitting all previously generated (simultaneous) packets. We complete the proof.
\end{IEEEproof}

\section{Proof of Theorem~\ref{thm:UB_long_sampling_delay_case}}\label{apx:proof_UB_long_sampling_delay_case}
\begin{IEEEproof}
We analyze PAoI violation probability for the long sampling period case as follows,
\begin{align}
\text{Pr}(A_{i}(k) \geq x) \overset{(a)}{\leq} &\text{Pr}\left( b + T^+_{i}(k-1) + V_{i}(k) \geq x \right) \nonumber \\
= &P_r\left( T^+_{i}(k-1) + V_{i}(k) \geq (x - b) \right) \nonumber \\
\overset{(b)}{\leq} &\mathbb{E}\left[ e^{\theta (T^+_{i}(k-1) + V_{i}(k))} \right] e^{-\theta(x - b)} \label{eqn:prob contains event E}  \\
\overset{(c)}{=} &\sum_{\ell=0}^{n-1} e^{(\ell+2)\Lambda(\theta)} \text{Pr}\left( E_{i,\ell}(k) \right) \cdot e^{-\theta(x - b)} \label{eqn:prob with summation term},
\end{align}
where (a) applies Lemma \ref{lemma:peak age long}, (b) uses the Chernoff bound with a constant $\theta>0$. In (c), we denote $\ell$ as the number of transmissions between two successive
transmissions for source $i$ and denote $E_{i,\ell}(k)$ as the event that there exist $\ell$ transmission before transmitting update $(i,k)$. Next, we keep on derive \eqref{eqn:prob with summation term}, 
\begin{align}
    \eqref{eqn:prob with summation term} &\overset{(d)}{=}= e^{-\theta(x - b)} \sum_{\ell=0}^{n-1} e^{(\ell+2)\Lambda(\theta)}   \cdot \left( \sum_{\mathbf{y}_{i,\ell}\in \mathcal{S}_{i,\ell}} g\left( \mathbf{y}_{i,\ell}, n, \mathbbm{1}, \boldsymbol{\mu} \right) \frac{\mu_i}{1-\sum_{j\in \mathbf{y}_{i,\ell}} \mu_j} \right) \nonumber\\
    &\overset{(e)}{\leq} \exp\left(- \theta (x-b)\right) \ n \! \!\max_{0\leq \ell \leq n-1} \left\{ \exp \left( \vphantom{\max_{0\leq \ell \leq n-1}} f_i(\ell,n,\boldsymbol{\mu}) \right) \right\} \nonumber\\
    &= \exp \left\{ -\theta x + \theta b + n\!\max_{0\leq \ell \leq n-1} f_i(\ell,n, \boldsymbol{\mu})  \right\}, \label{eqn:violation prob with f}
\end{align}
where (d) applies the multivariate Wallenius noncentral hypergeometric distribution to represent $\ell$ transmissions under a specific $\mathbf{y}_{i,\ell}$ and after $\ell$ transmissions, the probability of transmitting update $(i,k)$ can be represented as $\frac{\mu_i}{1-\sum_{j\in \mathbf{y}_{i,\ell}} \mu_j}$. (e) applies the union bound and
\begin{align}
    &f_i(\ell,n, \boldsymbol{\mu}) = (\ell+2) \Lambda(\theta) + \log \left( \sum_{\mathbf{y}_{i,\ell}\in S_{i,\ell}} g(\mathbf{y}_{i,\ell}, n, \mathbbm{1}, \boldsymbol{\mu}) \frac{\mu_i}{1-\sum_{j\in \mathbf{y}_{i,\ell}} \mu_j} \right). 
\end{align}
Since \eqref{eqn:violation prob with f} holds for every $\theta$, we choose the best one,
\begin{align}
    &\text{Pr}(A_{i}(k) \geq x) \leq \exp\left( -\sup_{\theta>0} \left\{ \theta x - \theta b - n\!\max_{0\leq \ell \leq n-1}f_i(\ell,n, \boldsymbol{\mu}) \right\} \right), 
\end{align}
and we complete the proof.
\end{IEEEproof}

\section{Proof of Corollary~\ref{coro:asym_decay_rate_rand_policy_long_sampling_delay}}\label{apx:proof_asymptotic_decay_rate_rand_policy_long_sampling_delay}

\begin{IEEEproof}
We first apply Lemma~\ref{lemma:exponent term converge to n}, which implies,
\begin{equation}
     f_i(n-1,n,\boldsymbol{\mu}) = \max_{0\leq \ell \leq n-1} f_i(\ell,n,\boldsymbol{\mu}),
\end{equation}
as $n$ is sufficiently large. Let the sampling period be defined as $b = n b'$ and the threshold as $x = n x'$, both of which scale with $n$. We then begin by taking the logarithm of the PAoI violation probability, dividing it by $n$, and letting the number of sources $n$ go to infinity,
\begin{align}
&\lim_{n \to \infty} \frac{1}{n} \log \text{Pr} (A_i(k) \geq nx') \nonumber \\ 
\overset{(a)}{\leq} &\lim_{n \to \infty} \frac{1}{n} \log \left( n \cdot \exp \left\{ -\theta x' + \theta b' + \!\max_{0\leq \ell \leq n-1} f_i(\ell,n, \boldsymbol{\mu})  \right\} \right) \nonumber\\
= &\lim_{n \to \infty} \frac{1}{n} \left\{ \log n - \theta x' + \theta b' + \!\max_{0\leq \ell \leq n-1} f_i(\ell,n, \boldsymbol{\mu})  \right\} \nonumber\\
\overset{(b)}{=} &- \left( \theta x' - \theta b' - f_i(\infty,\infty,\boldsymbol{\mu}) \right),
\end{align}
where (a) applies \eqref{eqn:violation prob with f} and (b) applies Lemma \ref{lemma:exponent term converge to n}. Then, we choose the best $\theta$ and complete the proof by the following equation,
\begin{align}
    &-\lim_{n \to \infty} \frac{1}{n} \log P_r \left( A_{g,i}{(k)} \geq nx' \right) \geq \sup_{\theta>0} \left\{ \theta x' - \theta b' - f_i(\infty,\infty,\boldsymbol{\mu}) \right\}. 
\end{align}
\end{IEEEproof}

\section{Proof of Lemma~\ref{lemma:peak age short}}\label{apx:proof UB PAoI short sampling}
\begin{IEEEproof}
    We start from \eqref{eqn:A(k)} and substitute $D_{i}(k)$ and $W_{i}(k)$ with \eqref{eqn:D(k)} and \eqref{eqn:W(k)}, respectively,
    \begin{align}
        &A_{i}(k) = D_{i}(k) - S_{i}(k-1)\nonumber\\
        &\overset{(a)}{=} S_{i}(k) + W_{i}(k-1) + T_{i}(k-1) + N_{i}(k-1) - (I_{i}(k-1)+1) b + V_{i}(k) -S_{i}(k-1) \nonumber\\
        &\overset{(b)}{=} W_{i}(k-1) + T_{i}(k-1) + N_{i}(k-1) + V_{i}(k) \nonumber\\
        &\overset{(c)}{\leq} b + T_i(k-1) + V_i(k). \nonumber
    \end{align}
    where (a) applies \eqref{eqn:D(k)} and \eqref{eqn:W(k)}; (b) is due to $S_i(k) - S_i(k-1)=(I_i(k-1)+1)b$, which includes the inter-arrival times of all preempted packets between $S_i(k-1)$ and $S_i(k)$, as well as the time between the last preempted packet and the arrival of update $(i,k)$. (c) applies $W_i(k-1) \leq b$ and $N_i(k-1)=0$ due to the assumption that BS is always busy. We then complete the proof.
\end{IEEEproof}

\section{Proof of Theorem~\ref{thm:UB_short_sampling_delay_case}}\label{apx:proof_UB_short_sampling_delay_case}
\begin{IEEEproof}
For the short sampling period case, we assume that all queues remain non-empty after each update. We will now begin analyzing the PAoI violation probability as presented in \eqref{eqn:prob contains event E} in Appendix \ref{apx:proof_UB_long_sampling_delay_case},
\begin{align}
\eqref{eqn:prob contains event E}
&\overset{(a)}{\leq}  \sum_{\ell=0}^{\infty} e^{ (\ell+1) \Lambda(\theta)}  \left( (1 - \mu_i)^{\ell} \mu_i \right)  e^{-n\theta x}  e^{n\theta b} \nonumber\\
&\overset{(b)}{=} \frac{e^{\Lambda(\theta)} \cdot \mu_i}{1 - e^{\Lambda(\theta)}(1 - \mu_i)}  e^{-n\theta x}  e^{n\theta b},
\end{align}
where (a) applies the geometric distribution and (b) holds if $\Lambda(\theta) < \log \left( \frac{1}{1 - \mu_i} \right)$. Moreover, we choose a specific $\theta$ that provides the tightest upper bound.
\begin{align}
    \text{Pr}\left(A_{i}(k) \geq nx\right) \leq \inf_{\theta>0} \left\{ e^{-n\theta (x-b)} \frac{e^{\Lambda(\theta)}\mu_i}{1 - e^{\Lambda(\theta)}(1 - \mu_i)} \right\}.
\end{align}
\end{IEEEproof}

\section{Proof of Corollary~\ref{coro:asym_decay_rate_rand_policy_short_sampling_delay}}\label{apx:proof_asymptotic_decay_rate_rand_policy_short_sampling_delay}
\begin{IEEEproof}
To get the asymptotic decay rate, we take the logarithm of the PAoI violation probability, divide it by $n$, and let the number of sources go to infinity. We have,
\begin{align}
&\lim_{n \to \infty} \frac{1}{n} \log \text{Pr} ( A_i(k) \geq nx' ) \vphantom{\frac{e^{\Lambda(\theta)} \cdot \mu_i}{1 - e^{\Lambda(\theta)} (1 - \mu_i)}} \nonumber\\
\leq &\lim_{n \to \infty} \frac{1}{n} \log \left( \inf_{\theta>0} \left\{ \frac{e^{\Lambda(\theta)} \cdot \mu_i}{1 - e^{\Lambda(\theta)} (1 - \mu_i)} \cdot e^{-n\theta x'} \cdot e^{n\theta b'} \right\} \right) \nonumber \\
= &\lim_{n \to \infty} \frac{1}{n} \inf_{\theta>0} \left\{ \Lambda(\theta) + \log \mu_i - \log \left( 1 - e^{\Lambda(\theta)} (1 - \mu_i) \right)  - n \theta x' + n \theta b' \right\} \vphantom{\frac{e^{\Lambda(\theta)} \cdot \mu_i}{1 - e^{\Lambda(\theta)} (1 - \mu_i)}}\nonumber \\
= &\inf_{\theta>0} \left\{ -\theta x' + \theta b' \right\}. \vphantom{\frac{e^{\Lambda(\theta)} \cdot \mu_i}{1 - e^{\Lambda(\theta)} (1 - \mu_i)}} 
\end{align}
\end{IEEEproof}

\section{Proof of Lemma \ref{lemma:exponent term converge to n}}\label{apx:proof_exp_term_increasing}
To prove that for sufficiently large $n$, for all $0\leq j \leq n-1$, as $n$ is sufficiently large, we have the following equation,
\begin{equation}
    f_i(n-1,n,\boldsymbol{\mu}) \geq f_i(j,n,\boldsymbol{\mu}).
\end{equation}
We first prove that for sufficiently large $n$, for all $0 \leq \ell \leq n-2$, the following equation holds,
\begin{equation}\label{eqn:exp_term_increasing}
    e^{(\ell+1) \Lambda(\theta)} \cdot \text{Pr}\left(E_{i, \ell}(k)\right) \leq e^{(\ell+2) \Lambda(\theta)} \cdot \text{Pr}\left(E_{i, \ell+1}(k)\right). 
\end{equation} 
We begin with the definition of the two probabilities and do some algebra,
\begin{align}
\text{Pr}\left(\text{E}_{i, \ell}(k)\right) &= \sum_{\mathbf{y}_{i,\ell} \in S_{i,\ell}} g\left(\mathbf{y}_{i,\ell}, n, \mathbbm{1}, \boldsymbol{\mu} \right) \cdot \frac{\mu_i}{1-\sum_{j\in \mathbf{y}_{i,\ell}} \mu_j} \label{eqn:prob event ell},\\
\text{Pr}\left(\text{E}_{i, \ell+1}(k)\right) &= \sum_{\mathbf{y}_{i,\ell+1} \in S_{i,\ell+1}} g\left(\mathbf{y}_{i,\ell+1}, n, \mathbbm{1}, \boldsymbol{\mu} \right) \cdot \frac{\mu_i}{1-\sum_{j\in \mathbf{y}_{i,\ell+1}} \mu_j} \nonumber\\
&= \sum_{\mathbf{y}_{i,\ell} \in S_{i,\ell}} g\left(\mathbf{y}_{i,\ell}, n, \mathbbm{1}, \boldsymbol{\mu} \right) \cdot \sum_{t \in \mathcal{T}'_{i,\ell}} \frac{\mu_t}{1-\sum_{j \in \mathbf{y}_{i,\ell}}\mu_j} \frac{\mu_i}{1-\sum_{j\in \mathbf{y}_{i,\ell}} \mu_j -\mu_t} \nonumber\\
&\overset{(a)}{\geq} \left( \sum_{\mathbf{y}_{i,\ell} \in S_{i,\ell}} g\left(\mathbf{y}_{i,\ell}, n, \mathbbm{1}, \boldsymbol{\mu} \right) \frac{\mu_i}{1-\sum_{j\in \mathbf{y}_{i,\ell}} \mu_j}  \right) \frac{(n-\ell-1) \mu_{\text{min}} }{ 1-\ell \cdot \mu_{\text{min}}}, \label{eqn:prob event ell plus 1}
\end{align}
where $\mathcal{T}'_{i,\ell}=\{j\lvert \mathbf{y}_{i,\ell,j}=0, j \neq i\}$ is the set contains all sources that have not been served except $i$, (a) lower bound the weight $\mu_j$ and $\mu_t$ as $\mu_{\text{min}}$, which $\mu_{\text{min}}$ is the smallest assigned weight. Next, we divide the two probabilities, we have,
\begin{align}\label{eqn:prob ratio upper bound n to infinte}
    \frac{\text{Pr}\left(\text{E}_{i, \ell}(k)\right)}{\text{Pr}\left(\text{E}_{i, \ell+1}(k)\right)} \leq \frac{\eqref{eqn:prob event ell}}{\eqref{eqn:prob event ell plus 1}} = \frac{1-\ell \cdot \mu_{\text{min}}}{(n-\ell-1)\mu_{\text{min}})}.
\end{align}
Let the number of sources $n$ goes to infinite,
\begin{equation}
    \lim_{n\to\infty} \frac{1-\ell \cdot \mu_{\text{min}}}{(n-\ell-1)\mu_{\text{min}})} \to 0. 
\end{equation}
By \eqref{eqn:prob ratio upper bound n to infinte}, we then further get,
\begin{equation}
    \lim_{n\to\infty} \frac{\text{Pr}\left(\text{E}_{i, \ell}(k)\right)}{\text{Pr}\left(\text{E}_{i, \ell+1}(k)\right)} \leq \lim_{n\to\infty} \frac{1-\ell \cdot \mu_{\text{min}}}{(n-\ell-1)\mu_{\text{min}})} \leq \epsilon \leq e^{\Lambda(\theta)},
\end{equation}
where $\epsilon$ is a small value. Since \eqref{eqn:exp_term_increasing} holds, we know that the terms of $f_i(\ell,n,\boldsymbol{\mu})$ monotonically increases as $\ell$ increase. This implies $\argmax_{0\leq \ell \leq n-1} f_i(\ell,n,\boldsymbol{\mu}) = n-1$ and we complete the proof.

\end{appendices}

\end{document}